\documentclass[journal=jacsat,manuscript=article]{achemso}\setkeys{acs}{keywords = true}

\usepackage{amsmath}
\usepackage{amssymb}
\usepackage{chemformula} 
\usepackage{graphicx}
\usepackage{dcolumn}
\usepackage{xcolor}
\usepackage{bm}
\usepackage{hyperref}
\usepackage[mathlines]{lineno}
\usepackage[T1]{fontenc} 

\author{David Abella}
\affiliation[1]{Instituto de F\'{\i}sica Interdisciplinar y Sistemas Complejos IFISC (CSIC-UIB),\\ Campus UIB, 07122 Palma de Mallorca, Spain.}

\author{Giancarlo Franzese}
\email{gfranzese@ub.edu}
\affiliation{Secci\'o de F\'isica Estad\'istica i Interdisciplin\`aria - Departament de F\'isica de la Mat\`eria Condensada, Universitat de Barcelona, Mart\'{\i} i Franqu\`es 1, 08028 Barcelona, Spain.}
\alsoaffiliation{Institut de Nanoci\`{e}ncia i Nanotecnologia, Universitat de Barcelona, 08028 Barcelona, Spain.}

\author{Javier Hern\'andez-Rojas}
\email{jhrojas@ull.edu.es}
\affiliation{Departamento de F\'isica e IUdEA, Universidad de La Laguna, 38205 La Laguna, Tenerife, Spain.}

\keywords{Water, Many-body, coordination shell, molecular dynamics, interaction radius, dipole interaction}

\title{Many-Body Contributions in Water Nano-Clusters}

\begin{document}

\begin{abstract}

  {\centering
    \includegraphics[width=0.8\columnwidth]{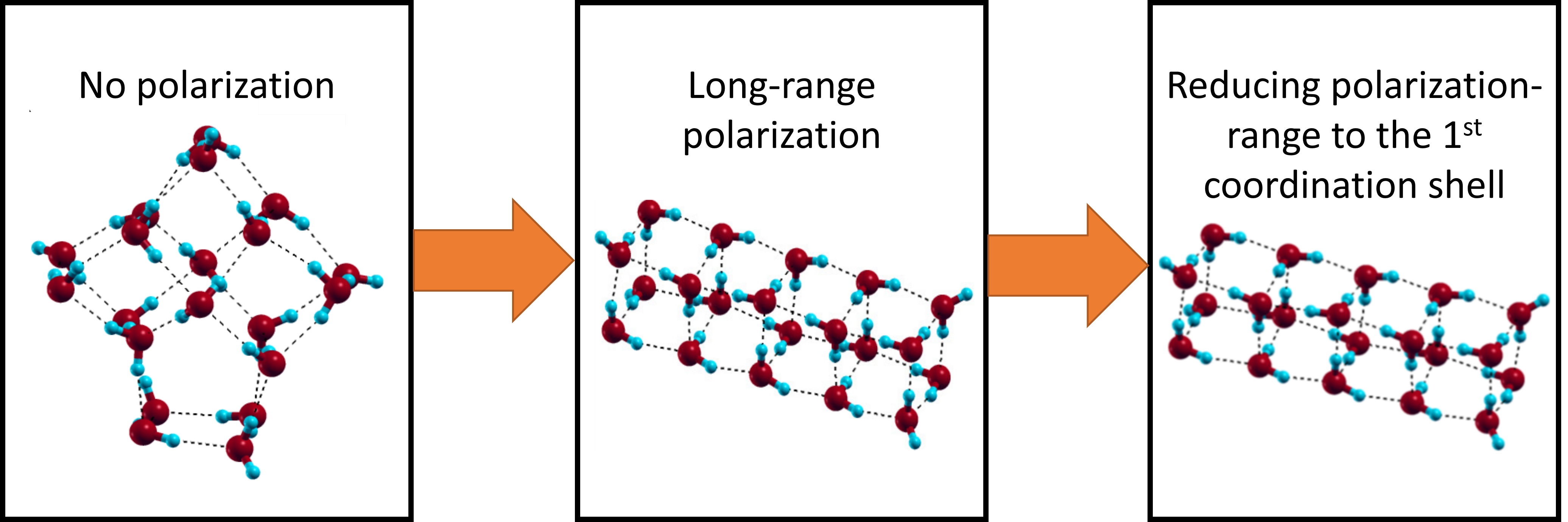}
    \par
  }
  Many-body interactions in water are known to be important but difficult to treat in atomistic models and often are included only as a correction. Polarizable models treat them explicitly, with long-range many-body potentials, within their classical approximation. However, their calculation is computationally expensive. Here, we evaluate how relevant the contributions to the many-body interaction associated with different coordination shells are. We calculate the global energy minimum, and the corresponding configuration, for nano-clusters of up to 20 water molecules. We find that including the first coordination shell, i.e., the five-body term of the central molecule, is enough to approximate within 5\% the global energy minimum and its structure. We show that this result is valid for three different polarizable models, the Dang-Chang, the MB-pol, and the Kozack-Jordan potentials. This result suggests a strategy to develop many-body potentials for water that are reliable and, at the same time, computationally efficient.
\end{abstract}

\section{Introduction}

Water is an object of intense research for its unusual properties and central role in many areas of science and technology \cite{Franks-2000}. In the last 50 years, computer simulations have contributed to understanding some of these peculiar phenomena. On one hand, {\it ab-initio} calculations have been used to predict structural \cite{BurnhamX02,FanourgakisAX04,XantheasA04,Xantheas2019} and dynamical properties \cite{Silvestrelli-1999,Phillip-2001,Chen-2003,Kuo-2004,Phillip-2001} of water from quantum calculations. However, this technique requires a high computational cost and generally treats small systems of the order of $\approx 100$ molecules. On the other hand, classical simulations can be useful for understanding these behaviors and can also deal with bigger systems involving thousands of atoms or molecules \cite{guvencac95,guvenca96,yehm99,yoshiiymo98, EgorovBL02,tainter:104304}.

However, in classical Molecular Dynamics or Monte Carlo simulations, the choice of the force field is critical. Many of the classical force fields for water are based on pairwise dispersion-repulsion and electrostatic interactions, and the many-body contributions are neglected. One of the most popular potential models is the TIP4P \cite{jorgensen-1983} and its family of TIP4P-like models \cite{Vega2005, Abascal:2005bh}. In these models, each water molecule is considered rigid, with four sites, including oxygen, two hydrogens, and one, often called the {\it M} site, located along the bisector of the oxygen-hydrogen vectors. The water-water interaction is given by Lennard-Jones and Coulomb pairwise potentials. The parameters of the TIP4P potential are chosen to replicate the structural properties of bulk water at standard temperature and pressure. However, this nonpolarizable potential cannot reproduce, for example, high-density properties where the many-body interactions play a fundamental role \cite{paricaud2005}. Thus, the polarizable models are built to overcome these deficiencies and are based on explicitly incorporating a non-additive term.  
  
In this work, we aim to elucidate how relevant are the many-body effects on the energetic and structural properties of water nano-clusters. To achieve this goal, we first employ the rigid-body polarizable Dang-Chang (DC) potential \cite{dang-1997}. This model, defined in the Methods section, is characterized by two terms. One is associated with the pairwise additive and the other with the non-additive polarization term. 

To describe the importance of the many-body effects in water, we introduce a cut-off radius in the polarization term and, employing the Basin-Hopping global optimization technique \cite{wales-1997}, we identify the putative global energy minimum for several selected water clusters with nm-size. 
We expect to obtain global minimum structures similar to the known TIP4P configurations for a minimal value of the cut-off radius, 
whereas, for a larger cut-off, the DC minimum structures. 
We ask if we can find an intermediate cut-off value that could reproduce the DC results within a reasonable approximation.

\section{Results}

\subsection{Energy dependence}

\begin{figure*}
    \centering
	\includegraphics[width=\columnwidth]{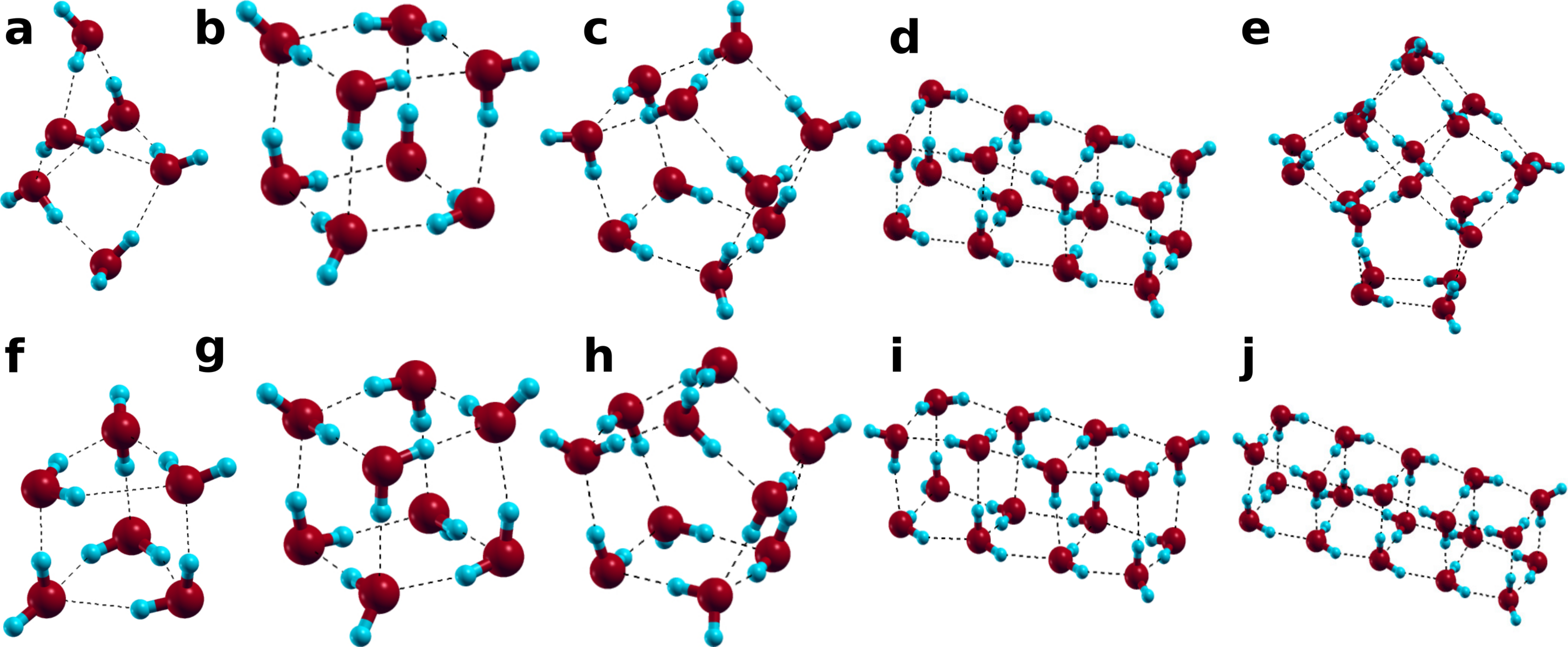}
	\caption{The lowest-energy configuration for a cluster with 6 (a,f), 8 (b, g), 10 (c,h), 16 (d, i), and 20 (e, j) water molecules calculated for the model with the shortest cut-off 1~\text{\AA},  as an approximation (see text) of the TIP4P-like model (a - e), and with the largest cut-off 20~\text{\AA}, as in the DC limit (f - j). For $N = 16$, our method recovers the same water molecule orientations in the two limits.}
	\label{vmd}
\end{figure*}

Based on previous work \cite{hernandez-rojas-2010}, we focus on water nano-clusters with 6 to 20 molecules (Fig.~\ref{vmd}). We minimize the energy of each cluster, as described in the Methods section, by applying a cut-off {\it exclusively} to the non-additive polarization term of the DC potential.

Surprisingly, we observe that the cut-off also influences the Lennard-Jones and the Coulomb energy contributions as an {\it indirect} effect due to the structural changes of the global-energy minima induced by the polarization.

\begin{figure}
	\includegraphics[width=0.6\columnwidth]{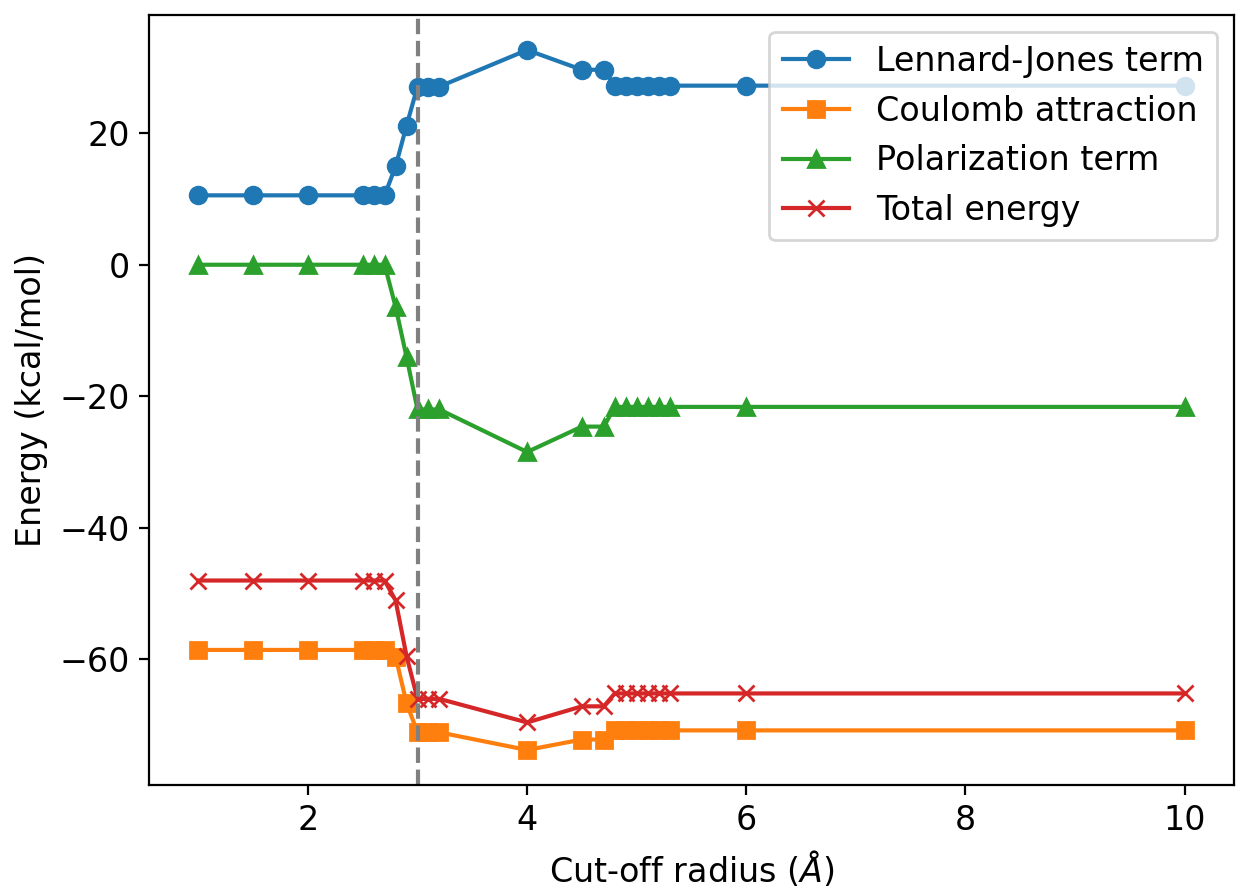} 
	\caption{Contributions to binding energy for the minimum-energy configuration as a function of the cut-off radius $r$ for a water cluster of 8 molecules (octamer). The Lennard-Jones contribution (blue circles) is always repulsive, while Coulomb (orange squares) and polarization (green triangles) terms are attractive.  The Coulomb contribution is three times larger than the other two terms, dominating the total energy (red crosses). The three contributions show a clear correlation and a non-monotonic dependence on the cut-off radius, with a significant drop at $r = 3$ \AA~ (gray dashed line). The minimum-energy configurations for the shortest and the largest cut-off are shown in Fig.\ref{vmd} b and g, respectively.} 
	\label{min_8}
\end{figure}

We find that, for all cluster sizes, the resulting binding energy for the minimum-energy configurations is non-monotonic as a function of the cut-off radius $r$ (
Fig.1 and Fig.\ref{min_8} in the Supporting Information).

When $r$ is larger than $d_{\rm max}$, the largest O-O distance in the cluster, all the energy contributions must converge to a constant value. For the octamer,  for example, this is true at $r>5~\text{\AA}$ because it is $d_{\rm max} \simeq$4.85~$\text{\AA}$. Thus, we recover the energy calculated for the DC model \cite{hernandez-rojas-2010} and the corresponding configurations (Fig.\ref{vmd} f-j). 

When $r$ is shorter than the water first-coordination shell distance, $r \lesssim 3 \; \textrm{\AA}$, we expect that the polarization energy vanishes and the other terms are constant. 
Under these circumstances, although the DC parameters for the isotropic potentials are slightly different from those of the TIP4P, 
one could expect that the DC model approximates well the TIP4P-like's minimum energy configurations (Fig. \ref{vmd} a-e). We verify it is so for all the cases we considered except for $N=20$. However, there is no apparent change between the results for the two extreme cut-off radii for 16 molecules (Fig. \ref{vmd} d-i). We will discuss the surprising result for 16 molecules in a separate section.

\begin{figure}
	\includegraphics[width=0.6\columnwidth]{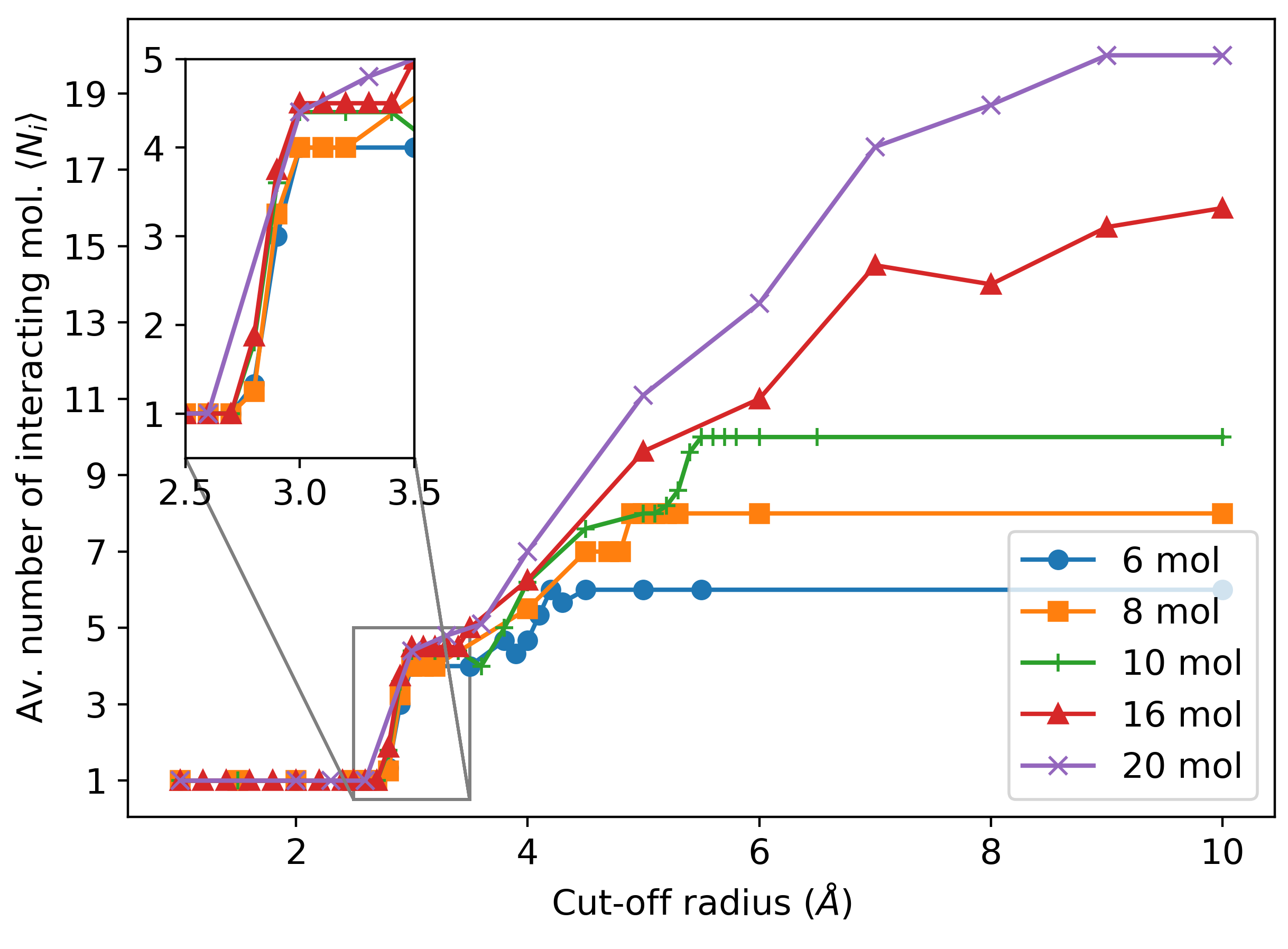}
	\caption{Average number $\langle N_i \rangle$ of molecules interacting via the many-body potential	for the global minimum configuration as a function of the cut-off radius $r$ for the different water cluster sizes.} 
	\label{N_6}
\end{figure}

Therefore, by increasing the cut-off radius, we tune the global-minimum energy from the unpolarizable to the polarizable model.
In particular, for all the cluster sizes, we observe a switching behavior in the binding energy when we cross the $r \simeq 3 \text{ \AA}$ threshold as a consequence of the sudden change in the number of water molecules interacting via the polarization potential. To illustrate this point, we calculate the average number $\langle N_i \rangle$ of interacting water molecules as a function of $r$ in each cluster (Fig.~\ref{N_6}). The average is over all the molecules of the same clusters.
Above $r \simeq 3~\text{\AA}$, the number $\langle N_i \rangle$ jumps from 1 to 4 or 5  
for the three smaller and the two larger clusters, respectively (in bulk water, the number of molecules within the first shell would be 5 in a tetrahedral configuration and 6 in a local high-density configuration with an interstitial molecule).

By further increasing $r$, $\langle N_i \rangle$ has step-like increases at each coordination-shell distance for the different clusters.  However, the steps smooth out for larger clusters due to the broadening of the O-O distances distribution over which we calculate $\langle N_i \rangle$. As a consequence of the variation of $\langle N_i \rangle$, the calculated binding energy has a non-monotonic behavior with $r$, e.g., with a minimum at  $r=4.0$~\text{\AA}, for the octamer (Fig.\ref{min_8}) or maxima for the other clusters (Fig.1 in the Supporting Information), due to partial contributions of the coordination shells, as we discuss next.

\subsection{Minimum-energy configurations}

The visual comparison of our minimum-energy configurations at the two extreme cut-off radii, $r \lesssim 3~\text{\AA}$ and  $r>d_{\rm max}$, with the configurations of the global-energy minima found in the literature for TIP4P \cite{Wales1998} and DC \cite{hernandez-rojas-2010},  confirms that, by tuning $r$, we move between the space of minima of the two limiting models.
For example, for the hexamer (6 molecules), we modulate between the cage structure of the TIP4P \cite{james-2005} for $r < 3 \text{\AA}$ (Fig.~\ref{vmd}a) to the trigonal prism of the DC \cite{hernandez-rojas-2010} for $r > d_{\rm max}=4.16~\text{\AA}$  
(Fig.~\ref{vmd}d).

\begin{figure*}
	\centering
	\includegraphics[width=\columnwidth]{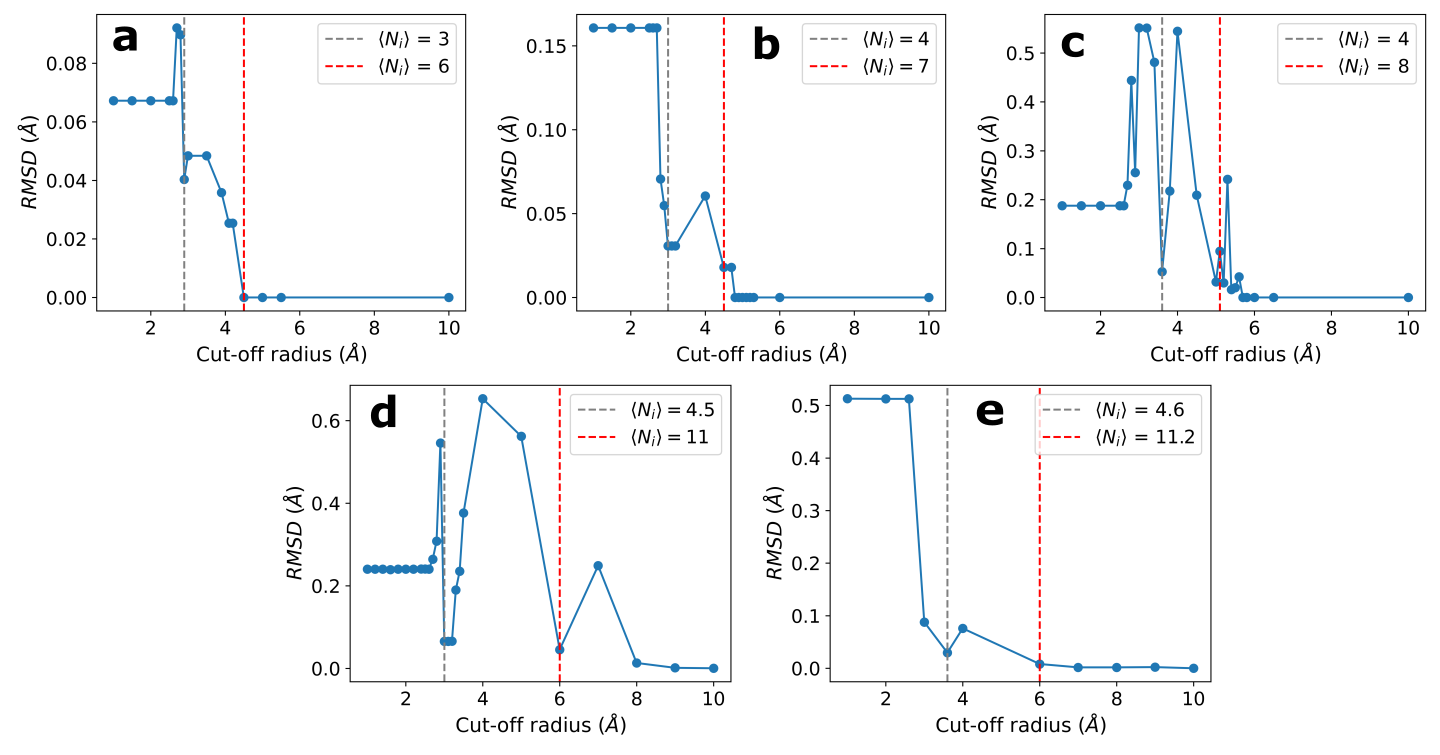} 
	\caption{Root mean square deviation (RMSD) between a cut-off cluster and the reference DC cluster as a function of the cut-off radius $r$ for (a) 6, (b) 8, (c) 10, (d) 16 and (e) 20 molecules. Vertical dashed lines mark the first value of the average number of interacting molecules $\langle N_i \rangle$ computed where the coordination shells are complete: gray for the first and red for the second coordination shells.} 
	\label{dev}
\end{figure*}

To make this comparison more quantitative, we calculate how close the configuration at a given cut-off $r$ (the {\it cut-off} cluster) is to the DC {\it reference} cluster with the long-range many-body contributions by computing the Root-mean-square deviation (RMSD) between the two configurations
\begin{equation}
    RMSD = \sqrt{ \frac{1}{N}\sum_{i}^{N} (r_i-R_i)^2 },
\end{equation}
%
where $R_i$ and $r_i$ are the O-O distances within the reference and the cut-off cluster, respectively. 
The index $i$ labels the different distances between molecules in the cluster, and $N$ is the size of the cluster.
We evaluate the RMSD over all the possible permutations of distances within the clusters and consider the minimum as our estimate (Fig. \ref{dev}). 

As expected, we recover the minimum-energy configuration of the full DC case for large cut-off values. Interestingly, we find that the dependence of the RMSD is not monotonic with the cut-off, displaying several minima and maxima. In particular, all the clusters have a minimum RMSD $< 0.05 \text{ \AA}$ at $r \simeq 3 \text{ \AA}$, i.e., the distance of the first coordination shell. Thus, the minimum energy configuration becomes very similar to the DC limit when we include the first coordination shell (see Fig. 3 in the Supporting Information to observe the minimum energy configurations for $r \sim 3$ \AA).

At cut-off radii that do not correspond to the distance of a coordination shell, the RMSD increases, i.e., the agreement between the configuration of minimum energy and the DC reference case is reduced. This happens although the total energy of the cluster reaches a global minimum, as for the octamer at $r=4.0$~\text{\AA}
 (Fig.~\ref{min_8}). Indeed, this minimum is a consequence of the many-body interaction acting on a number of molecules $\langle N_i \rangle$  that is intermediate between two consecutive coordination shells (Fig.~\ref{dev} b).

\begin{figure}[h]
	\centering
	\includegraphics[width=0.6\columnwidth]{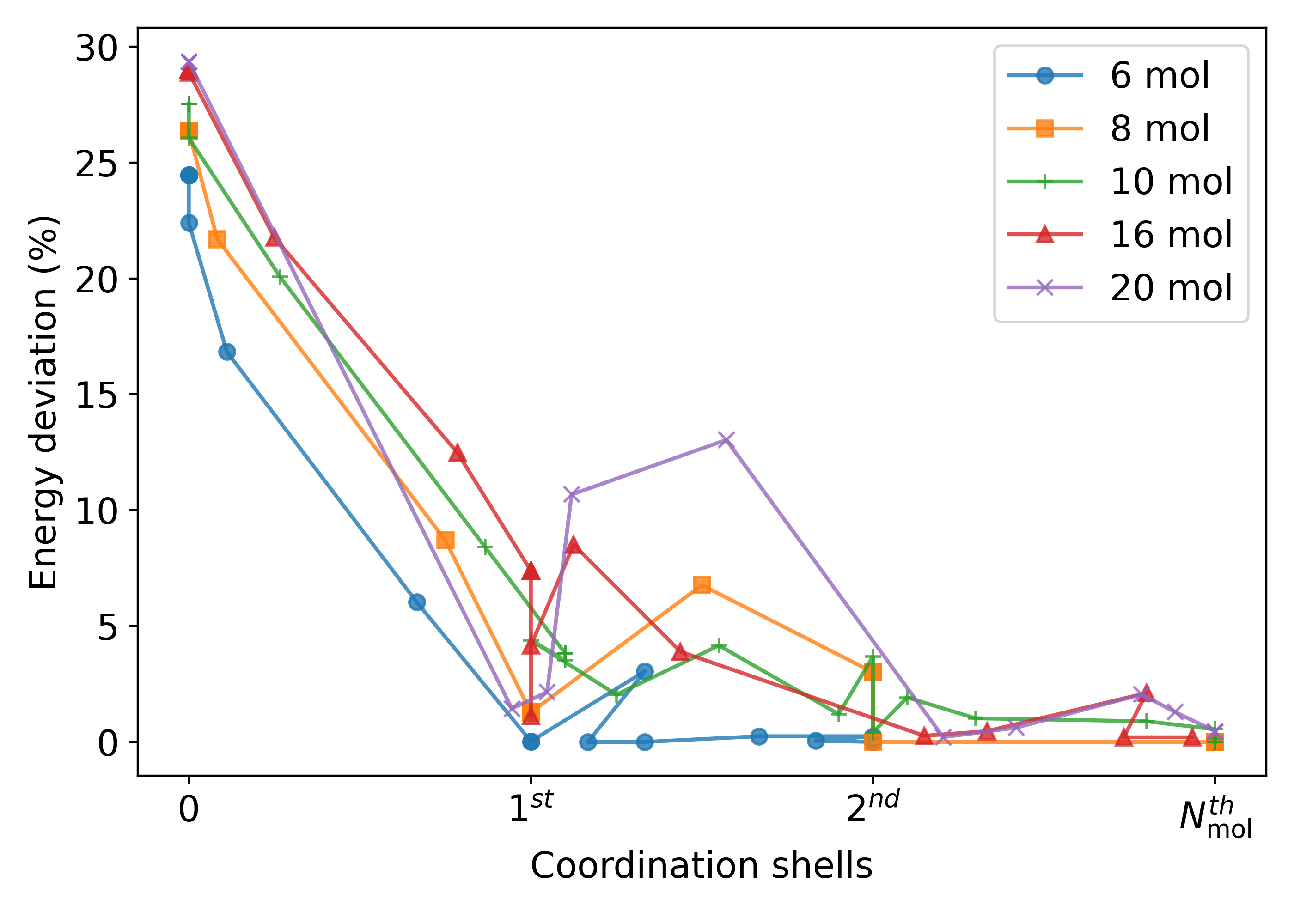} 
\caption{Parametric plot of the energy deviation, compared to the reference DC value,  as a function of coordination shell number for clusters with 6 (blue circles), 8 (orange squares), 10 (green pluses), 16 (red triangles) and 20 (purple crosses) molecules. The deviations are within 5\% when the first coordination shell is completed, i.e., when  $ \langle N_i \rangle \simeq 4$ and $r \simeq 3 \AA$ (see Fig. 2 in Supporting Information).} 
	\label{Et_N}
\end{figure}

Next, we study the total energy deviation from the DC reference results as a function of the cut-off radius $r$. 
We find that, for all cluster sizes, the energy deviations at the first coordination shell ($\simeq 3 \text{ \AA}$) are  $\lesssim 5\%$ than the reference DC energy (Fig. 2 in Supporting Information). The drop in the energy deviation at the first coordination-shell distance is evident when we represent it as a function of $\langle N_i \rangle$, observing that for our clusters, the first coordination shell includes from 3 to 4.6 molecules (Fig. \ref{Et_N}). 

Furthermore,  the energy deviation increases for larger cut-off radii that are intermediate between two consecutive coordination shells, and it does not improve significantly when we include more shells.
Overall, we conclude that both the RMSD and the energy deviation of the cut-off cluster drop below 5\% compared with the  DC reference cluster when the cut-off coincides with the first coordination shell. To prove that these results are not model dependent, we perform the same analysis using the MB-pol potential \cite{babin2013development,medders2014development,babin2014development}, which has been shown to correctly predict the properties of water across a wide range of thermodynamic conditions \cite{medders2015infrared,reddy2016accuracy}, and the Kozack-Jordan (KJ) potential \cite{KJ}. In the Supporting Information, we show that we find the same behavior as a function of the cut-off radius, proving the generality of our result.

\section{The nano-cluster with 16 molecules} 

The RMSD is an accurate observable to differentiate similar clusters. For example, for 8 and 10 molecules, the clusters with the cut-off at the first coordination shell are similar to the reference DC clusters. However, they have different ordering in the hydrogen bond directions. Consequently, the RMSD of the corresponding clusters is finite (Fig. \ref{dev} b, and c).

The accuracy of RMSD allows us to understand also the case with 16 molecules, displaying minimum-energy configurations with no polarization contribution  (Fig. \ref{vmd} d) and full polarization contribution (Fig. \ref{vmd} i) that are indistinguishable by the naked eye. 
We find this surprising result also for 12 molecules (not shown) that likewise minimize their energy by clustering as {\it fused} cubes. Nevertheless, although the configurations of fused cubes look the same in both short and large cut-off limits, they are not. 

Indeed, the RMSD for 16 molecules at the first coordination shell is $>1$\% larger than the reference DC cluster (Fig. \ref{dev} d). A more refined analysis (not presented here) reveals that these differences are due to minimal variations in the O-O distances of the short-cut-off cluster that account for an energy deviation $>6$\% larger than the reference DC cluster (Fig. 2 in the Supporting Information).

\section{Conclusions}

We analyze the effect of introducing a cut-off $r$ in three many-body, long-range, polarizable potentials: the DC model, which adds polarizability to a TIP4P-like potential, the MB-pol, and the KJ potential. First, we consider DC-water nano-clusters of up to 20 molecules and calculate their minimum-energy configuration. We evaluate the root-mean-square deviation of the structure and the deviation of the energy of the minima compared to the reference unrestricted simulations with the full DC polarizable potential \cite{hernandez-rojas-2010}. We repeat the calculations for the MB-pol and the KJ potentials. To minimize the computational cost, we 1) replicate the entire procedure only for two representative cluster sizes (8 and 20) of the KJ water, and 2) use the minimum-energy configurations of DC-water as a starting point for the MB-pol analysis.  

Surprisingly, we find that the cut-off, although applied only to the polarization term of the potential and not the other interactions, induces variations in the energies of each term of the potential. This is a consequence of the change in the global minimum structure dominated by the many-body interaction.

For the DC model, we find that the deviations are not monotonic with the cut-off radius $r$. When $r$ does not correspond to the average distance of a coordination shell, the nano-clusters reach artificial minima with energy below the full DC case but with a significant structural deviation from the correct minimum. When the $r$ corresponds to the first shell, we find an agreement within 5\% for the RMSD of the configuration and its total energy relative to the reference values.   
For the larger 1-nm clusters considered here, the first shell comprises five molecules. 
The same behavior is found for the MB-pol and KJ models. 
Furthermore, the same number of first coordinated molecules is also characteristic of the bulk water in a local tetrahedral structure.  

Therefore, our results show that approximating the long-range, many-body (polarization) interaction with a short-range, five-body interaction is better than using fewer-body terms, what may lead to improvements over recent potential energy functions that account for explicit short-range interactions only up to three-body contribution \cite{Cisneros:2016aa}.
The idea of representing the many-body effects in water via a five-body term has been explored in previous work \cite{gonzalez-2005}, and 
has been pursued in 
a coarse-grained model of water that preserves the molecular description of the hydrogen-bonds \cite{franzese-2002, FS2007, coronas2020franzesestanley}, successfully comparing with experiments  \cite{Mazza2011}, and allowing a better understanding of hydration water in protein physics \cite{Bianco:2015aa, BiancoPRX2017, Bianco-Navarro2019, Bianco:2020aa}. 
These approaches find support in our present results. 
It is likely that contribution beyond the first coordination shell 
 would be necessary to determine the correct minimum binding energy and other properties of water. However, we have shown that, for three different polarizable water models at the cost of (less than) 5\% error in interaction energies and the right global minimum structure, it is enough to consider just the contribution of the first coordination shell. 
This approximation would vastly reduce the computational cost of large-scale simulations of hydrated systems, including an effective approximation of the long-range many-body interactions.
 
\section{Methods}
 
Pairwise-additive Lennard-Jones and Coulomb terms plus a many-body polarization contribution describe the rigid-body polarization DC potential~\cite{dang-1997}. The Lennard-Jones term is applied between oxygen atoms, whereas the Coulomb interaction is on partial charges on the hydrogen and M sites. The polarization term is characterized by the isotropic molecular polarizability on the M site and the induced dipole moments due to the electric field produced by fixed charges in the system.

The putative global energy minima of water clusters were located using the Basin-Hopping method \cite{wales-1997}. This technique has been used successfully in atomic and molecular clusters \cite{james-2005,hr2006,hr2012,wales2003}. To treat the rigid-body orientational degrees of freedom, we employed the angle-axis scheme \cite{wales2005}. The advantage of this coordinate system is that it does not suffer from the ``gimbal lock'' problem, which can occur with the use of Euler angles \cite{wawak1992}. 

We perform four independent trajectories of 1$\times10^5$ basin-hopping steps for each cluster, starting with random geometries and a constant optimization temperature of $k_B T\simeq$ 3~kcal/mol. We attempt blocks of 100 translational and 200 angular moves with an acceptance ratio of 20$\%$.

To evaluate how the global energy minimum changes with the cut-off radius,   
first, we calculate the global minimum for 20 water molecules with a cut-off radius of 20.0~\text{\AA}. We check that this distance is enough to account for all many-body energy contributions for the cluster sizes considered here. Under this condition, we recover the global minimum obtained with the Dang-Chang Model \cite{hernandez-rojas-2010}. This preliminary analysis allows us to calculate the maximum O-O distance in each cluster, corresponding to the minimum cut-off radius needed to include the total contribution of the many-body potential.

Finally, for each cut-off radius $r$, we find the minimum energy configurations with the Basin-Hopping method and calculate the average, over all the cluster molecules, of the number $N_i$ of molecules interacting.
Since the global energy structure can change with the cut-off radius, the relative distance between the molecules varies, resulting in a non-monotonic $\langle N_i \rangle$ function, highlighting intriguing features of the global energy minimum configurations.

\begin{acknowledgement}

We would like to acknowledge Prof. Francesco Paesani and Dr. Marc Riera for the fruitful discussions about their MB-pol potential and also for their work modifying the MBX library for carrying out our cut-off radius analysis. D.A. acknowledges support from the Maria de Maeztu program MDM-2017-0711 of the MCIN/AEI/10.13039/501100011033. G.F. acknowledges support from the Spanish grants PGC2018-099277-B-C22 and PID2021-124297NB-C31, funded by MCIN/AEI/ 10.13039/ 501100011033 and ``ERDF A way of making Europe", and the ICREA Foundation (ICREA Academia prize). 
J.H.-R. acknowledges financial support from the Spanish ``Ministerio de Ciencia e Innovaci\'on'' through Grant PID2019-105225GB-I00 (MICINN/FEDER,UE).

\end{acknowledgement}

\begin{suppinfo}

The following files are available free of charge.
\begin{itemize}
  \item Supporting Information: Additional numerical results of the energy contributions for all the cluster sizes considered (Fig.1) and energy deviation dependence with the cut-off radius and the average number of interacting molecules (Fig.2).

  \item Minimum energy configurations at the 1st coordination shell (Fig. 3).

  \item Results using the MB-pol potential (Fig. 4).

  \item Results using the KJ potential (Fig. 5).
  
\end{itemize}

\end{suppinfo}

\section{Associated content}
The authors declare no competing financial interest. 
A preliminary version of this manuscript has been previously submitted to a pre-print server: 
D. Abella; G. Franzese; and J. Hern\'andez-Rojas. 
Many Body contributions of Water Nano-clusters. 
2022, 
Article number: arXiv:2206. 13976. 
Name of Repository: arXiv.
https://arxiv.org/abs/2206.13976 
(accessed June 28, 2022).

\bibliography{bibliography}

\providecommand{\latin}[1]{#1}
\providecommand*\mcitethebibliography{\thebibliography}
\csname @ifundefined\endcsname{endmcitethebibliography}
  {\let\endmcitethebibliography\endthebibliography}{}
\begin{mcitethebibliography}{46}
\providecommand*\natexlab[1]{#1}
\providecommand*\mciteSetBstSublistMode[1]{}
\providecommand*\mciteSetBstMaxWidthForm[2]{}
\providecommand*\mciteBstWouldAddEndPuncttrue
  {\def\EndOfBibitem{\unskip.}}
\providecommand*\mciteBstWouldAddEndPunctfalse
  {\let\EndOfBibitem\relax}
\providecommand*\mciteSetBstMidEndSepPunct[3]{}
\providecommand*\mciteSetBstSublistLabelBeginEnd[3]{}
\providecommand*\EndOfBibitem{}
\mciteSetBstSublistMode{f}
\mciteSetBstMaxWidthForm{subitem}{(\alph{mcitesubitemcount})}
\mciteSetBstSublistLabelBeginEnd
  {\mcitemaxwidthsubitemform\space}
  {\relax}
  {\relax}

\bibitem[Franks(2000)]{Franks-2000}
Franks,~F. \emph{Water}; RSC Paperbacks; The Royal Society of Chemistry:
  Cambridge, 2000; pp X001--X004\relax
\mciteBstWouldAddEndPuncttrue
\mciteSetBstMidEndSepPunct{\mcitedefaultmidpunct}
{\mcitedefaultendpunct}{\mcitedefaultseppunct}\relax
\EndOfBibitem
\bibitem[Burnham and Xantheas(2002)Burnham, and Xantheas]{BurnhamX02}
Burnham,~C.~J.; Xantheas,~S.~S. \emph{J. Chem. Phys.} \textbf{2002},
  \emph{116}, 1500--1510\relax
\mciteBstWouldAddEndPuncttrue
\mciteSetBstMidEndSepPunct{\mcitedefaultmidpunct}
{\mcitedefaultendpunct}{\mcitedefaultseppunct}\relax
\EndOfBibitem
\bibitem[Fanourgakis \latin{et~al.}(2004)Fanourgakis, Apra, and
  Xantheas]{FanourgakisAX04}
Fanourgakis,~G.~S.; Apra,~E.; Xantheas,~S.~S. \emph{J. Chem. Phys.}
  \textbf{2004}, \emph{121}, 2655--2663\relax
\mciteBstWouldAddEndPuncttrue
\mciteSetBstMidEndSepPunct{\mcitedefaultmidpunct}
{\mcitedefaultendpunct}{\mcitedefaultseppunct}\relax
\EndOfBibitem
\bibitem[Xantheas and Apra(2004)Xantheas, and Apra]{XantheasA04}
Xantheas,~S.~S.; Apra,~E. \emph{J. Chem. Phys.} \textbf{2004}, \emph{120},
  823--828\relax
\mciteBstWouldAddEndPuncttrue
\mciteSetBstMidEndSepPunct{\mcitedefaultmidpunct}
{\mcitedefaultendpunct}{\mcitedefaultseppunct}\relax
\EndOfBibitem
\bibitem[Rakshit \latin{et~al.}(2019)Rakshit, Bandyopadhyay, Heindel, and
  Xantheas]{Xantheas2019}
Rakshit,~A.; Bandyopadhyay,~P.; Heindel,~J.~P.; Xantheas,~S.~S. \emph{The
  Journal of Chemical Physics} \textbf{2019}, \emph{151}, 214307\relax
\mciteBstWouldAddEndPuncttrue
\mciteSetBstMidEndSepPunct{\mcitedefaultmidpunct}
{\mcitedefaultendpunct}{\mcitedefaultseppunct}\relax
\EndOfBibitem
\bibitem[Silvestrelli and Parrinello(1999)Silvestrelli, and
  Parrinello]{Silvestrelli-1999}
Silvestrelli,~P.~L.; Parrinello,~M. \emph{Phys. Rev. Lett.} \textbf{1999},
  \emph{82}, 3308--3311\relax
\mciteBstWouldAddEndPuncttrue
\mciteSetBstMidEndSepPunct{\mcitedefaultmidpunct}
{\mcitedefaultendpunct}{\mcitedefaultseppunct}\relax
\EndOfBibitem
\bibitem[Geissler \latin{et~al.}(2001)Geissler, Dellago, Chandler, Hutter, and
  Parrinello]{Phillip-2001}
Geissler,~P.~L.; Dellago,~C.; Chandler,~D.; Hutter,~J.; Parrinello,~M.
  \emph{Science} \textbf{2001}, \emph{291}, 2121--2124\relax
\mciteBstWouldAddEndPuncttrue
\mciteSetBstMidEndSepPunct{\mcitedefaultmidpunct}
{\mcitedefaultendpunct}{\mcitedefaultseppunct}\relax
\EndOfBibitem
\bibitem[Chen \latin{et~al.}(2003)Chen, Ivanov, Klein, and
  Parrinello]{Chen-2003}
Chen,~B.; Ivanov,~I.; Klein,~M.~L.; Parrinello,~M. \emph{Phys. Rev. Lett.}
  \textbf{2003}, \emph{91}, 215503\relax
\mciteBstWouldAddEndPuncttrue
\mciteSetBstMidEndSepPunct{\mcitedefaultmidpunct}
{\mcitedefaultendpunct}{\mcitedefaultseppunct}\relax
\EndOfBibitem
\bibitem[Kuo \latin{et~al.}(2004)Kuo, Mundy, McGrath, Siepmann, VandeVondele,
  Sprik, Hutter, Chen, Klein, Mohamed, Krack, and Parrinello]{Kuo-2004}
Kuo,~I.-F.~W.; Mundy,~C.~J.; McGrath,~M.~J.; Siepmann,~J.~I.; VandeVondele,~J.;
  Sprik,~M.; Hutter,~J.; Chen,~B.; Klein,~M.~L.; Mohamed,~F.; Krack,~M.;
  Parrinello,~M. \emph{The Journal of Physical Chemistry B} \textbf{2004},
  \emph{108}, 12990--12998\relax
\mciteBstWouldAddEndPuncttrue
\mciteSetBstMidEndSepPunct{\mcitedefaultmidpunct}
{\mcitedefaultendpunct}{\mcitedefaultseppunct}\relax
\EndOfBibitem
\bibitem[G\"uven\c{c} and Choi(1995)G\"uven\c{c}, and Choi]{guvencac95}
G\"uven\c{c},~M.~A.,~Z. B.~Anderson; Choi,~B.~H. \emph{Z. Phys. D}
  \textbf{1995}, \emph{35}, 51--55\relax
\mciteBstWouldAddEndPuncttrue
\mciteSetBstMidEndSepPunct{\mcitedefaultmidpunct}
{\mcitedefaultendpunct}{\mcitedefaultseppunct}\relax
\EndOfBibitem
\bibitem[G\"uven\c{c} and Anderson(1996)G\"uven\c{c}, and Anderson]{guvenca96}
G\"uven\c{c},~Z.~B.; Anderson,~M.~A. \emph{Z. Phys. D} \textbf{1996},
  \emph{36}, 171--183\relax
\mciteBstWouldAddEndPuncttrue
\mciteSetBstMidEndSepPunct{\mcitedefaultmidpunct}
{\mcitedefaultendpunct}{\mcitedefaultseppunct}\relax
\EndOfBibitem
\bibitem[Yeh and Mou(1999)Yeh, and Mou]{yehm99}
Yeh,~Y.~L.; Mou,~C.~Y. \emph{J. Phys. Chem. B} \textbf{1999}, \emph{103},
  3699--3705\relax
\mciteBstWouldAddEndPuncttrue
\mciteSetBstMidEndSepPunct{\mcitedefaultmidpunct}
{\mcitedefaultendpunct}{\mcitedefaultseppunct}\relax
\EndOfBibitem
\bibitem[Yoshii \latin{et~al.}(1998)Yoshii, Yoshie, Miura, and
  Okazaki]{yoshiiymo98}
Yoshii,~N.; Yoshie,~H.; Miura,~S.; Okazaki,~S. \emph{J. Chem. Phys.}
  \textbf{1998}, \emph{109}, 4873--4884\relax
\mciteBstWouldAddEndPuncttrue
\mciteSetBstMidEndSepPunct{\mcitedefaultmidpunct}
{\mcitedefaultendpunct}{\mcitedefaultseppunct}\relax
\EndOfBibitem
\bibitem[Egorov \latin{et~al.}(2002)Egorov, Brodskaya, and
  Laaksonen]{EgorovBL02}
Egorov,~A.~V.; Brodskaya,~E.~N.; Laaksonen,~A. \emph{Mol. Phys.} \textbf{2002},
  \emph{100}, 941--952\relax
\mciteBstWouldAddEndPuncttrue
\mciteSetBstMidEndSepPunct{\mcitedefaultmidpunct}
{\mcitedefaultendpunct}{\mcitedefaultseppunct}\relax
\EndOfBibitem
\bibitem[Tainter and Skinner(2012)Tainter, and Skinner]{tainter:104304}
Tainter,~C.~J.; Skinner,~J.~L. \emph{J. Chem. Phys.} \textbf{2012}, \emph{137},
  104304\relax
\mciteBstWouldAddEndPuncttrue
\mciteSetBstMidEndSepPunct{\mcitedefaultmidpunct}
{\mcitedefaultendpunct}{\mcitedefaultseppunct}\relax
\EndOfBibitem
\bibitem[Jorgensen \latin{et~al.}(1983)Jorgensen, Chandrasekhar, Madura, Impey,
  and Klein]{jorgensen-1983}
Jorgensen,~W.~L.; Chandrasekhar,~J.; Madura,~J.~D.; Impey,~R.~W.; Klein,~M.~L.
  \emph{The Journal of Chemical Physics} \textbf{1983}, \emph{79},
  926--935\relax
\mciteBstWouldAddEndPuncttrue
\mciteSetBstMidEndSepPunct{\mcitedefaultmidpunct}
{\mcitedefaultendpunct}{\mcitedefaultseppunct}\relax
\EndOfBibitem
\bibitem[Vega and Abascal(2005)Vega, and Abascal]{Vega2005}
Vega,~C.; Abascal,~J. L.~F. \emph{Journal of Chemical Physics} \textbf{2005},
  \emph{123}, 144504\relax
\mciteBstWouldAddEndPuncttrue
\mciteSetBstMidEndSepPunct{\mcitedefaultmidpunct}
{\mcitedefaultendpunct}{\mcitedefaultseppunct}\relax
\EndOfBibitem
\bibitem[Abascal and Vega(2005)Abascal, and Vega]{Abascal:2005bh}
Abascal,~J. L.~F.; Vega,~C. \emph{The Journal of Chemical Physics}
  \textbf{2005}, \emph{123}, 234505--12\relax
\mciteBstWouldAddEndPuncttrue
\mciteSetBstMidEndSepPunct{\mcitedefaultmidpunct}
{\mcitedefaultendpunct}{\mcitedefaultseppunct}\relax
\EndOfBibitem
\bibitem[Paricaud \latin{et~al.}(2005)Paricaud, P{\v r}edota, Chialvo, and
  Cummings]{paricaud2005}
Paricaud,~P.; P{\v r}edota,~M.; Chialvo,~A.~A.; Cummings,~P.~T. \emph{The
  Journal of Chemical Physics} \textbf{2005}, \emph{122}, 244511\relax
\mciteBstWouldAddEndPuncttrue
\mciteSetBstMidEndSepPunct{\mcitedefaultmidpunct}
{\mcitedefaultendpunct}{\mcitedefaultseppunct}\relax
\EndOfBibitem
\bibitem[Dang and Chang(1997)Dang, and Chang]{dang-1997}
Dang,~L.~X.; Chang,~T.-M. \emph{The Journal of Chemical Physics} \textbf{1997},
  \emph{106}, 8149--8159\relax
\mciteBstWouldAddEndPuncttrue
\mciteSetBstMidEndSepPunct{\mcitedefaultmidpunct}
{\mcitedefaultendpunct}{\mcitedefaultseppunct}\relax
\EndOfBibitem
\bibitem[Wales and Doye(1997)Wales, and Doye]{wales-1997}
Wales,~D.~J.; Doye,~J. P.~K. \emph{The Journal of Physical Chemistry A}
  \textbf{1997}, \emph{101}, 5111--5116\relax
\mciteBstWouldAddEndPuncttrue
\mciteSetBstMidEndSepPunct{\mcitedefaultmidpunct}
{\mcitedefaultendpunct}{\mcitedefaultseppunct}\relax
\EndOfBibitem
\bibitem[Hern{\'a}ndez-Rojas \latin{et~al.}(2010)Hern{\'a}ndez-Rojas, Calvo,
  Rabilloud, Bret{\'o}n, and Gomez~Llorente]{hernandez-rojas-2010}
Hern{\'a}ndez-Rojas,~J.; Calvo,~F.; Rabilloud,~F.; Bret{\'o}n,~J.;
  Gomez~Llorente,~J.~M. \emph{The Journal of Physical Chemistry A}
  \textbf{2010}, \emph{114}, 7267--7274\relax
\mciteBstWouldAddEndPuncttrue
\mciteSetBstMidEndSepPunct{\mcitedefaultmidpunct}
{\mcitedefaultendpunct}{\mcitedefaultseppunct}\relax
\EndOfBibitem
\bibitem[Wales and Hodges(1998)Wales, and Hodges]{Wales1998}
Wales,~D.~J.; Hodges,~M.~P. \emph{Chemical Physics Letters} \textbf{1998},
  \emph{286}, 65--72\relax
\mciteBstWouldAddEndPuncttrue
\mciteSetBstMidEndSepPunct{\mcitedefaultmidpunct}
{\mcitedefaultendpunct}{\mcitedefaultseppunct}\relax
\EndOfBibitem
\bibitem[James \latin{et~al.}(2005)James, Wales, and
  Hern{\'a}ndez-Rojas]{james-2005}
James,~T.; Wales,~D.~J.; Hern{\'a}ndez-Rojas,~J. \emph{Chemical Physics
  Letters} \textbf{2005}, \emph{415}, 302--307\relax
\mciteBstWouldAddEndPuncttrue
\mciteSetBstMidEndSepPunct{\mcitedefaultmidpunct}
{\mcitedefaultendpunct}{\mcitedefaultseppunct}\relax
\EndOfBibitem
\bibitem[Babin \latin{et~al.}(2013)Babin, Leforestier, and
  Paesani]{babin2013development}
Babin,~V.; Leforestier,~C.; Paesani,~F. \emph{Journal of chemical theory and
  computation} \textbf{2013}, \emph{9}, 5395--5403\relax
\mciteBstWouldAddEndPuncttrue
\mciteSetBstMidEndSepPunct{\mcitedefaultmidpunct}
{\mcitedefaultendpunct}{\mcitedefaultseppunct}\relax
\EndOfBibitem
\bibitem[Medders \latin{et~al.}(2014)Medders, Babin, and
  Paesani]{medders2014development}
Medders,~G.~R.; Babin,~V.; Paesani,~F. \emph{Journal of chemical theory and
  computation} \textbf{2014}, \emph{10}, 2906--2910\relax
\mciteBstWouldAddEndPuncttrue
\mciteSetBstMidEndSepPunct{\mcitedefaultmidpunct}
{\mcitedefaultendpunct}{\mcitedefaultseppunct}\relax
\EndOfBibitem
\bibitem[Babin \latin{et~al.}(2014)Babin, Medders, and
  Paesani]{babin2014development}
Babin,~V.; Medders,~G.~R.; Paesani,~F. \emph{Journal of chemical theory and
  computation} \textbf{2014}, \emph{10}, 1599--1607\relax
\mciteBstWouldAddEndPuncttrue
\mciteSetBstMidEndSepPunct{\mcitedefaultmidpunct}
{\mcitedefaultendpunct}{\mcitedefaultseppunct}\relax
\EndOfBibitem
\bibitem[Medders and Paesani(2015)Medders, and Paesani]{medders2015infrared}
Medders,~G.~R.; Paesani,~F. \emph{Journal of Chemical Theory and Computation}
  \textbf{2015}, \emph{11}, 1145--1154\relax
\mciteBstWouldAddEndPuncttrue
\mciteSetBstMidEndSepPunct{\mcitedefaultmidpunct}
{\mcitedefaultendpunct}{\mcitedefaultseppunct}\relax
\EndOfBibitem
\bibitem[Reddy \latin{et~al.}(2016)Reddy, Straight, Bajaj, Huy~Pham, Riera,
  Moberg, Morales, Knight, G{\"o}tz, and Paesani]{reddy2016accuracy}
Reddy,~S.~K.; Straight,~S.~C.; Bajaj,~P.; Huy~Pham,~C.; Riera,~M.;
  Moberg,~D.~R.; Morales,~M.~A.; Knight,~C.; G{\"o}tz,~A.~W.; Paesani,~F.
  \emph{The Journal of chemical physics} \textbf{2016}, \emph{145},
  194504\relax
\mciteBstWouldAddEndPuncttrue
\mciteSetBstMidEndSepPunct{\mcitedefaultmidpunct}
{\mcitedefaultendpunct}{\mcitedefaultseppunct}\relax
\EndOfBibitem
\bibitem[Kozack and Jordan(1992)Kozack, and Jordan]{KJ}
Kozack,~R.~E.; Jordan,~P.~C. \emph{The Journal of Chemical Physics}
  \textbf{1992}, \emph{96}, 3120--3130\relax
\mciteBstWouldAddEndPuncttrue
\mciteSetBstMidEndSepPunct{\mcitedefaultmidpunct}
{\mcitedefaultendpunct}{\mcitedefaultseppunct}\relax
\EndOfBibitem
\bibitem[Cisneros \latin{et~al.}(2016)Cisneros, Wikfeldt, Ojam{\"a}e, Lu, Xu,
  Torabifard, Bart{\'o}k, Cs{\'a}nyi, Molinero, and Paesani]{Cisneros:2016aa}
Cisneros,~G.~A.; Wikfeldt,~K.~T.; Ojam{\"a}e,~L.; Lu,~J.; Xu,~Y.;
  Torabifard,~H.; Bart{\'o}k,~A.~P.; Cs{\'a}nyi,~G.; Molinero,~V.; Paesani,~F.
  \emph{Chemical Reviews} \textbf{2016}, \emph{116}, 7501--7528\relax
\mciteBstWouldAddEndPuncttrue
\mciteSetBstMidEndSepPunct{\mcitedefaultmidpunct}
{\mcitedefaultendpunct}{\mcitedefaultseppunct}\relax
\EndOfBibitem
\bibitem[Gonz{\'a}lez \latin{et~al.}(2005)Gonz{\'a}lez, Hern{\'a}ndez-Rojas,
  and Wales]{gonzalez-2005}
Gonz{\'a}lez,~B.~S.; Hern{\'a}ndez-Rojas,~J.; Wales,~D.~J. \emph{Chemical
  Physics Letters} \textbf{2005}, \emph{412}, 23--28\relax
\mciteBstWouldAddEndPuncttrue
\mciteSetBstMidEndSepPunct{\mcitedefaultmidpunct}
{\mcitedefaultendpunct}{\mcitedefaultseppunct}\relax
\EndOfBibitem
\bibitem[Franzese and Stanley(2002)Franzese, and Stanley]{franzese-2002}
Franzese,~G.; Stanley,~H.~E. \emph{Journal of Physics: Condensed Matter}
  \textbf{2002}, \emph{14}, 2201--2209\relax
\mciteBstWouldAddEndPuncttrue
\mciteSetBstMidEndSepPunct{\mcitedefaultmidpunct}
{\mcitedefaultendpunct}{\mcitedefaultseppunct}\relax
\EndOfBibitem
\bibitem[Franzese and Stanley(2007)Franzese, and Stanley]{FS2007}
Franzese,~G.; Stanley,~H.~E. \emph{Journal of Physics-Condensed Matter}
  \textbf{2007}, \emph{19}, 205126\relax
\mciteBstWouldAddEndPuncttrue
\mciteSetBstMidEndSepPunct{\mcitedefaultmidpunct}
{\mcitedefaultendpunct}{\mcitedefaultseppunct}\relax
\EndOfBibitem
\bibitem[Coronas \latin{et~al.}(2020)Coronas, Vilanova, Bianco, de~los Santos,
  and Franzese]{coronas2020franzesestanley}
Coronas,~L.~E.; Vilanova,~O.; Bianco,~V.; de~los Santos,~F.; Franzese,~G. In
  \emph{Properties of Water From Numerical and Experimental Perspectives};
  Martelli,~F., Ed.; CRC Press: Boca Raton, 2020; Chapter The Franzese-Stanley
  Coarse Grained Model for Hydration Water\relax
\mciteBstWouldAddEndPuncttrue
\mciteSetBstMidEndSepPunct{\mcitedefaultmidpunct}
{\mcitedefaultendpunct}{\mcitedefaultseppunct}\relax
\EndOfBibitem
\bibitem[Mazza \latin{et~al.}(2011)Mazza, Stokely, Pagnotta, Bruni, Stanley,
  and Franzese]{Mazza2011}
Mazza,~M.~G.; Stokely,~K.; Pagnotta,~S.~E.; Bruni,~F.; Stanley,~H.~E.;
  Franzese,~G. \emph{Proceedings of the National Academy of Sciences}
  \textbf{2011}, \emph{108}, 19873\relax
\mciteBstWouldAddEndPuncttrue
\mciteSetBstMidEndSepPunct{\mcitedefaultmidpunct}
{\mcitedefaultendpunct}{\mcitedefaultseppunct}\relax
\EndOfBibitem
\bibitem[Bianco and Franzese(2015)Bianco, and Franzese]{Bianco:2015aa}
Bianco,~V.; Franzese,~G. \emph{Physical Review Letters} \textbf{2015},
  \emph{115}, 108101--\relax
\mciteBstWouldAddEndPuncttrue
\mciteSetBstMidEndSepPunct{\mcitedefaultmidpunct}
{\mcitedefaultendpunct}{\mcitedefaultseppunct}\relax
\EndOfBibitem
\bibitem[Bianco \latin{et~al.}(2017)Bianco, Franzese, Dellago, and
  Coluzza]{BiancoPRX2017}
Bianco,~V.; Franzese,~G.; Dellago,~C.; Coluzza,~I. \emph{Phys. Rev. X}
  \textbf{2017}, \emph{7}, 021047\relax
\mciteBstWouldAddEndPuncttrue
\mciteSetBstMidEndSepPunct{\mcitedefaultmidpunct}
{\mcitedefaultendpunct}{\mcitedefaultseppunct}\relax
\EndOfBibitem
\bibitem[Bianco \latin{et~al.}(2019)Bianco, Alonso-Navarro, Di~Silvio, Moya,
  Cortajarena, and Coluzza]{Bianco-Navarro2019}
Bianco,~V.; Alonso-Navarro,~M.; Di~Silvio,~D.; Moya,~S.; Cortajarena,~A.~L.;
  Coluzza,~I. \emph{The Journal of Physical Chemistry Letters} \textbf{2019},
  \emph{10}, 4800--4804\relax
\mciteBstWouldAddEndPuncttrue
\mciteSetBstMidEndSepPunct{\mcitedefaultmidpunct}
{\mcitedefaultendpunct}{\mcitedefaultseppunct}\relax
\EndOfBibitem
\bibitem[Bianco \latin{et~al.}(2020)Bianco, Franzese, and
  Coluzza]{Bianco:2020aa}
Bianco,~V.; Franzese,~G.; Coluzza,~I. \emph{ChemPhysChem} \textbf{2020},
  \emph{21}, 377--384\relax
\mciteBstWouldAddEndPuncttrue
\mciteSetBstMidEndSepPunct{\mcitedefaultmidpunct}
{\mcitedefaultendpunct}{\mcitedefaultseppunct}\relax
\EndOfBibitem
\bibitem[Hern{\'a}ndez-Rojas \latin{et~al.}(2006)Hern{\'a}ndez-Rojas,
  Bret{\'o}n, Gomez~Llorente, and Wales]{hr2006}
Hern{\'a}ndez-Rojas,~J.; Bret{\'o}n,~J.; Gomez~Llorente,~J.~M.; Wales,~D.~J.
  \emph{The Journal of Physical Chemistry B} \textbf{2006}, \emph{110},
  13357--13362, PMID: 16821854\relax
\mciteBstWouldAddEndPuncttrue
\mciteSetBstMidEndSepPunct{\mcitedefaultmidpunct}
{\mcitedefaultendpunct}{\mcitedefaultseppunct}\relax
\EndOfBibitem
\bibitem[Hern{\'a}ndez-Rojas \latin{et~al.}(2012)Hern{\'a}ndez-Rojas, Calvo,
  Bret{\'o}n, and Gomez~Llorente]{hr2012}
Hern{\'a}ndez-Rojas,~J.; Calvo,~F.; Bret{\'o}n,~J.; Gomez~Llorente,~J.
  \emph{The Journal of Physical Chemistry C} \textbf{2012}, \emph{116},
  17019--17028\relax
\mciteBstWouldAddEndPuncttrue
\mciteSetBstMidEndSepPunct{\mcitedefaultmidpunct}
{\mcitedefaultendpunct}{\mcitedefaultseppunct}\relax
\EndOfBibitem
\bibitem[Wales(2003)]{wales2003}
Wales,~D.~J. \emph{Energy landscapes: Applications to clusters, biomolecules
  and glasses}; Cambridge University Press: Cambridge, 2003\relax
\mciteBstWouldAddEndPuncttrue
\mciteSetBstMidEndSepPunct{\mcitedefaultmidpunct}
{\mcitedefaultendpunct}{\mcitedefaultseppunct}\relax
\EndOfBibitem
\bibitem[Wales(2005)]{wales2005}
Wales,~D.~J. \emph{Philosophical Transactions of the Royal Society A:
  Mathematical, Physical and Engineering Sciences} \textbf{2005}, \emph{363},
  357--377\relax
\mciteBstWouldAddEndPuncttrue
\mciteSetBstMidEndSepPunct{\mcitedefaultmidpunct}
{\mcitedefaultendpunct}{\mcitedefaultseppunct}\relax
\EndOfBibitem
\bibitem[Wawak \latin{et~al.}(1992)Wawak, Wimmer, and Scheraga]{wawak1992}
Wawak,~R.~J.; Wimmer,~M.~M.; Scheraga,~H.~A. \emph{The Journal of Physical
  Chemistry} \textbf{1992}, \emph{96}, 5138--5145\relax
\mciteBstWouldAddEndPuncttrue
\mciteSetBstMidEndSepPunct{\mcitedefaultmidpunct}
{\mcitedefaultendpunct}{\mcitedefaultseppunct}\relax
\EndOfBibitem
\end{mcitethebibliography}

\end{document}